\documentclass[aps,prl,twocolumn,groupedaddress,showpacs]{revtex4}

\usepackage{graphicx}
\begin{document}

\title{Quantum nucleation in a single-chain magnet}

\author{W. Wernsdorfer$^1$, R. Cl\'erac$^2$, C. Coulon$^2$, L. Lecren$^2$, H. Miyasaka$^{3}$}


\affiliation{
$^1$Lab. L. N\'eel, associ\'e \`a l'UJF, CNRS, BP 166,
38042 Grenoble Cedex 9, France\\
$^2$Centre de Recherche Paul Pascal, CNRS UPR 8641, 115 Av. Dr. A.
Schweitzer, 33600 Pessac, France\\
$^3$Department of Chemistry, Graduate School of Science, Tokyo
Metropolitan University, Minami-Ohsawa 1-1, Hachioji, Tokyo 192-0397,Japan
}

\date{13 Sept. 2005}

\begin{abstract}
The field sweep rate $(v=dH/dt)$ and temperature $(T)$ dependence of the
magnetization reversal of a single-chain magnet (SCM)
is studied at low temperatures.
As expected for a thermally
activated process, the nucleation  field ($H_{\rm n}$) increases
with decreasing $T$ and increasing $v$. 
The set of $H_{\rm n}(T,v)$ data is analyzed
with a model of thermally activated nucleation
of magnetization reversal. Below 1~K, 
$H_{\rm n}$ becomes temperature independent but remains
strongly sweep rate dependent. 
In this temperature range, the reversal of the magnetization 
is induced by a quantum nucleation of a domain wall 
that then propagates due to the applied field.
\end{abstract}



\maketitle

Recent efforts in synthetic chemistry
has led to a quickly growing number of
magnetic systems that show slow relaxation
of magnetization. Apart of interests
for applications in, for instant,
ultrahigh density magnetic recording,
such systems are ideal to test 
theories. A well-known example is 
the single-molecule magnet (SMM)
that exhibit slow magnetization relaxation of their spin ground state, 
which is split by axial zero-field splitting~\cite{Sessoli93,Aubin96}. 
SMMs are among the most promising candidates for observing 
the limits between classical and quantum physics.
A more recent example is the single-chain magnet 
(SCM)~\cite{Caneschi01,Clerac02,Lescouezec03}
showing slow relaxation of magnetization
as the consequence of the uniaxial
anisotropy seen by each spin on the chain
and magnetic correlations between spins.
Although it seemed that there was a reasonable agreement
between the experimental data and Glauber's
theory of a 1D Ising spin chain~\cite{Glauber63},
it was shown that several other arguments should be
considered to fill the gap between the theory 
and the experimental results~\cite{Coulon04}.
The most important arguments concerned
the introduction of magnetic anisotropy
and finite-size effects.
Indeed, their influence on the static and dynamic 
properties of the SCMs was confirmed experimentally~\cite{Miyasaka03,Coulon04,Bogani04,Bogani05}.

Quantum tunneling of domain walls 
in a 1D mesoscopic ferromagnetic sample was theoretically
investigated~\cite{Stamp91,Chudnovsky92,Tatara94,Gunther94,Braun97,Shibata00} 
and crossover temperatures between
the classical and quantum regime were predicted. 
Domain wall nucleation and depinning were studied in single Ni wires
and showed indeed
a flattening of the temperature dependence of
the mean switching field ($H_{\rm sw}$) below about 
5~K~\cite{Hong95} and 1~K~\cite{WW_PRL96_Ni}.
Because of surface roughness and oxidation, 
the domain walls of a single wire were trapped 
at pinning centers. The pinning barrier 
decreases with an increase of the magnetic field. 
When the barrier is sufficiently small, 
thermally activated escape of the wall occurs. This 
is a stochastic process that can be characterized 
by a switching (depinning) field 
distribution. A flattening of the temperature 
dependence of $H_{\rm sw}$ and a 
saturation of the width of the switching field 
distribution were 
observed. The authors proposed that 
a domain wall escapes from its pinning site 
by thermal activation at high temperatures and by quantum tunneling 
below $T_{\rm c} \sim$ 5~K~\cite{Hong95} and 1~K~\cite{WW_PRL96_Ni}.
However, such crossover 
temperatures are about three 
orders of magnitude 
higher than the $T_{\rm c}$ predicted by current theories~\cite{Braun97}.
The propagation of a domain wall across an energy barrier in a domain
wall junction was also studied and preliminary investigations 
seems to indicate the possibility
of quantum tunneling below 0.7 K~\cite{Mangin97}.

In this letter, we show that tunneling can occur
in truly 1D systems like SCM provided that a driving field is
applied that lowers the energy barrier.
Indeed in zero applied field, the probability 
of tunneling is negligeable due to the exponential 
increase of the correlation length.

The studied SCM is a heterometallic
chain of Mn$^{III}$ and Ni$^{II}$ metal ions:
[Mn$_2$(saltmen)$_2$Ni(pao)$_2$(py)$_2$](ClO$_4$)$_2$
(saltmen$^{2-}$ = N,N'-(1,1,2,2-tetramethylethylene)
bis(salicylideneiminate); pao$^{-}$  = pyridine-2-aldoximate;
py = pyridine), called Mn$_2$Ni chain henceforth~\cite{Clerac02,Miyasaka03}.
At low temperatures, this compound can be described
as a chain of ferromagnetic coupled $S = 3$
[Mn$^{III}$-Ni$^{II}$-Mn$^{III}$] units. 

The spin system of the Mn$_2$Ni chain
can be described by an anisotropic Heisenberg model:
\begin{equation}
      \mathcal{H} = -J \sum_{i}\vec{S}_i\vec{S}_{i+1} -
                     D \sum_{i} S_{i,z}^2 - 
		     g\mu_{\rm B}\mu_0\sum_{i} \vec{S}_i\vec{H}
\label{Heisenberg}
\end{equation}
where $J$ is the ferromagnetic exchange constant
between the spin units and $D$ is the single-ion
anisotropy.
$D/k_{\rm B}$ = 2.5 K was obtained from magnetization measurements
as a function of a magnetic field applied perpendicular 
to the easy axis~\cite{Coulon04}. 
AC and DC relaxation time measurements showed
a unique relaxation time over 10 decades.
Above 2.7~K, the thermal dependence
of the relaxation time followed an Arrhenius
law with an activation energy of 74~K.
Below 2.7~K, a departure from this simple behavior was observed and a smaller
activation energy of 55~K was found around 2~K~\cite{Coulon04}.
The crossover at about 2.7~K
was interpreted as the manifestation of finite-size effects~\cite{Coulon04}.
Indeed, the activation energy of the relaxation 
time should decrease from $(4J + D)S^2$ to $(2J + D)S^2$ at the 
temperature where the correlation length equals the chain length~\cite{Coulon04}. 
The exchange energy $J/k_{\rm B} = 1.56 K$ was found which
was in agreement with independent thermodynamical measurements. 
A saturation of a semi-log plot of $\chi T$ versus $1/T$
was used to estimate the mean chain length of
about 100 Mn$_2$Ni units.

The relaxation rate at $H=0$ is extremely small below 1.4 K.
We applied therefore a magnetic field to study the low temperature
relaxation process.
The magnetization measurements were 
performed by using (i) a magnetometer consisting
of several 10~$\times$~10~$\mu$m$^{2}$ Hall-bars~\cite{Sorace03}
and (ii) an array of micro-SQUIDs~\cite{WW_ACP01}
on top of which a single crystal of 
Mn$_{2}$Ni was placed, for higher and lower fields
than 1.4 T, respectively.
The field can be applied in any direction by separately 
driving three orthogonal superconducting coils. 
The field was aligned with the easy axis of magnetization using
the transverse field method~\cite{WW_PRB04}.

\begin{figure}
\begin{center}
\includegraphics[width=.45\textwidth]{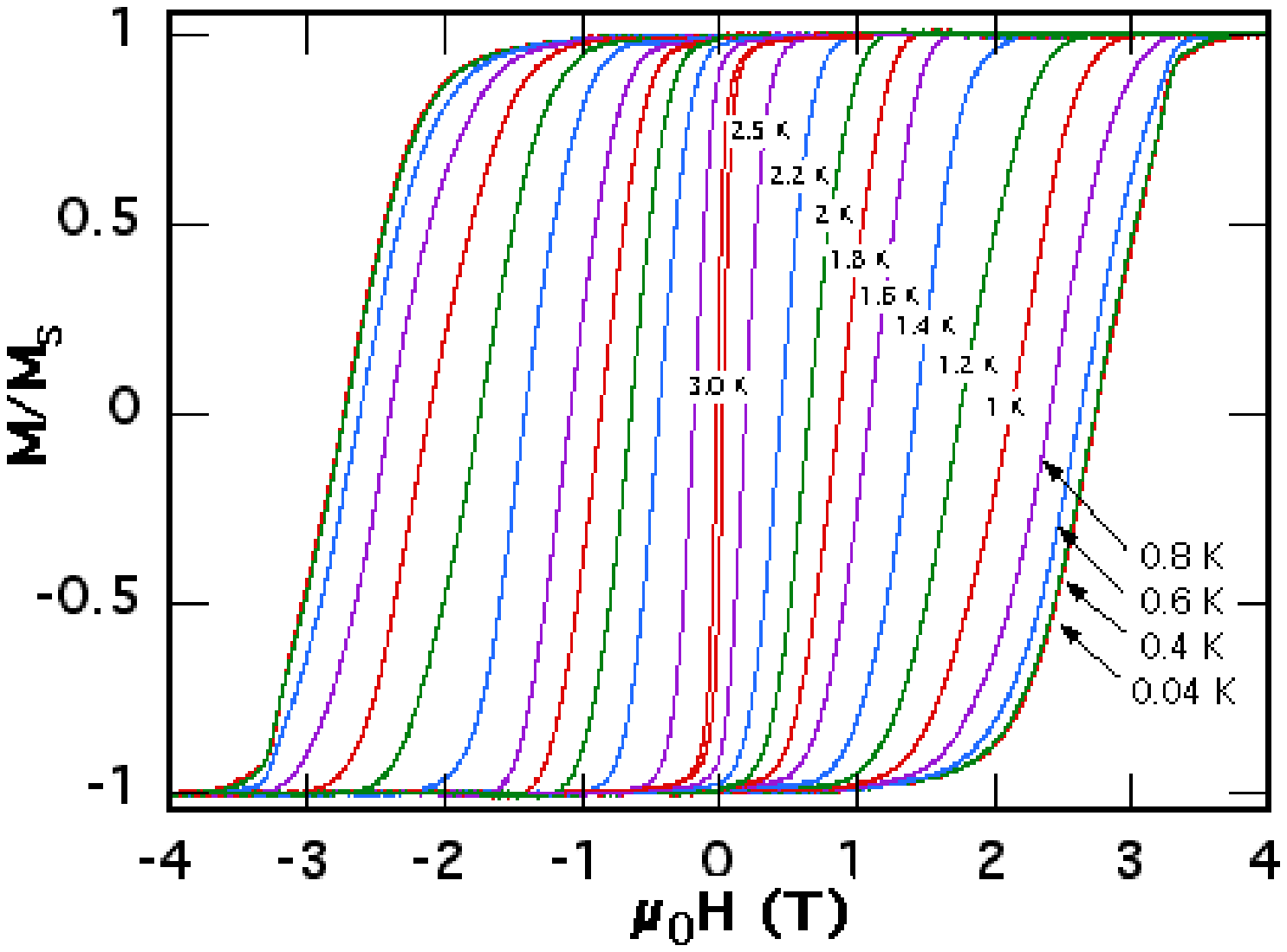}
\includegraphics[width=.45\textwidth]{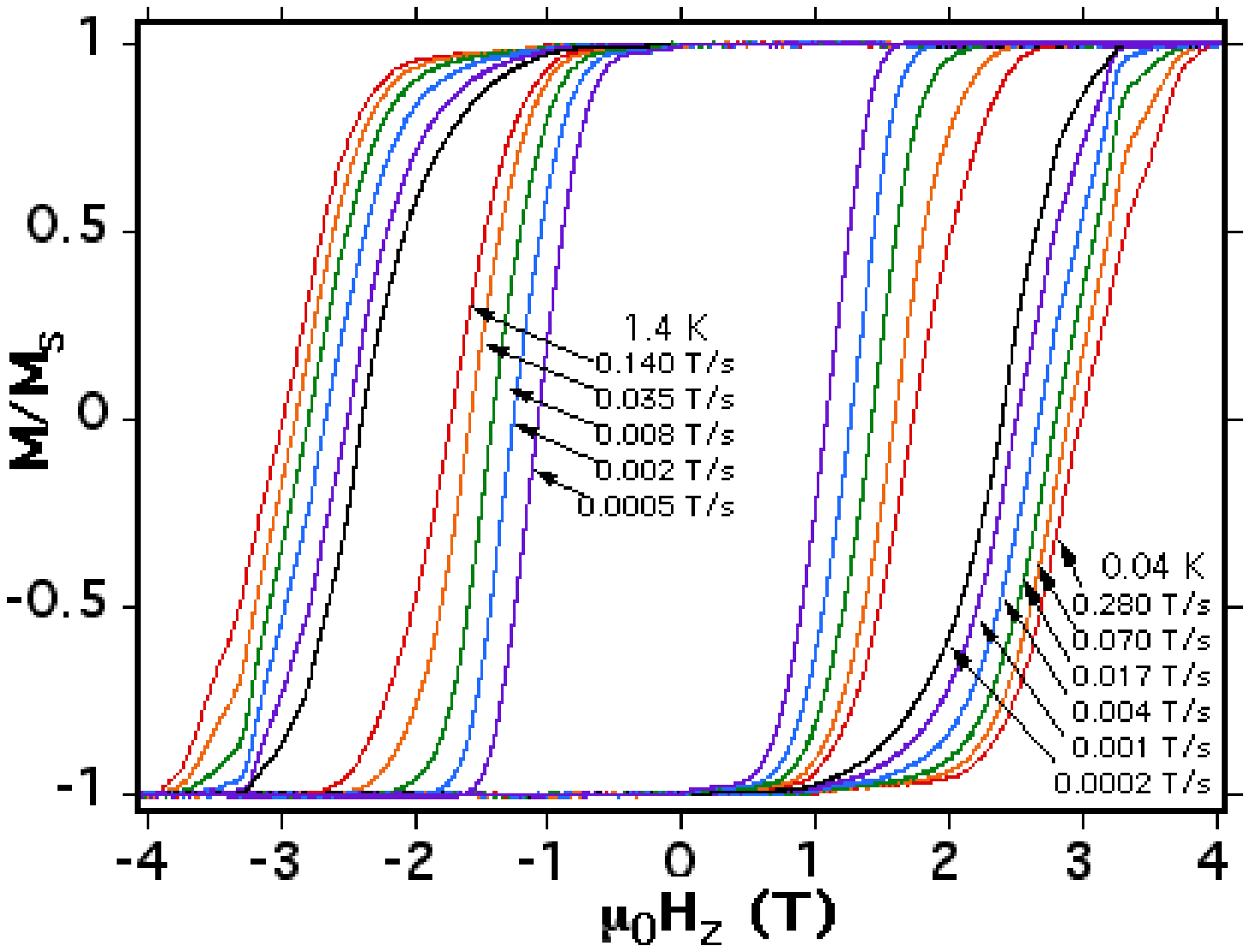}
\caption{(color online) Hysteresis loops for the Mn$_{2}$Ni 
chain at (a) several temperatures and 0.008 T/s
and (b) at 0.04 and 1.4 K and at several field sweep rates.}
\label{hyst_Mn2Ni}
\end{center}
\end{figure}

\begin{figure}
\begin{center}
\includegraphics[width=.45\textwidth]{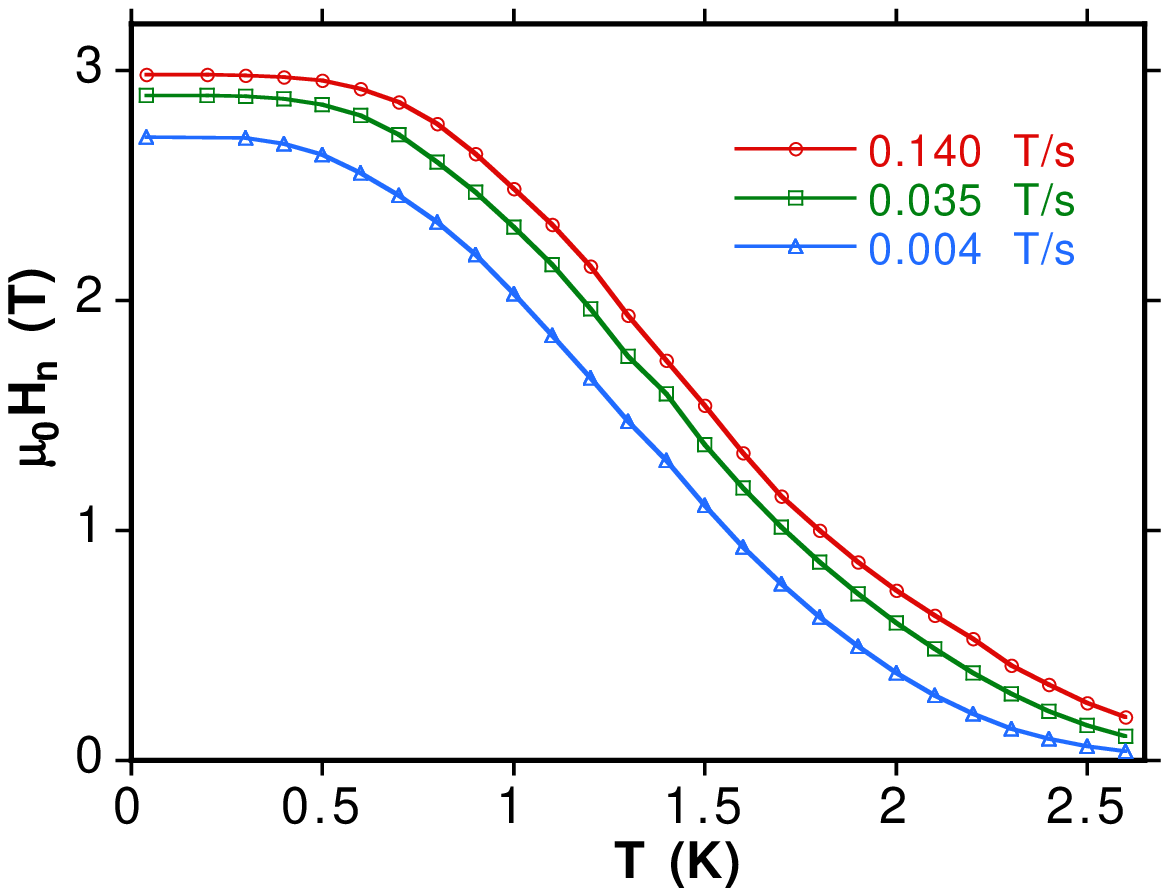}
\includegraphics[width=.45\textwidth]{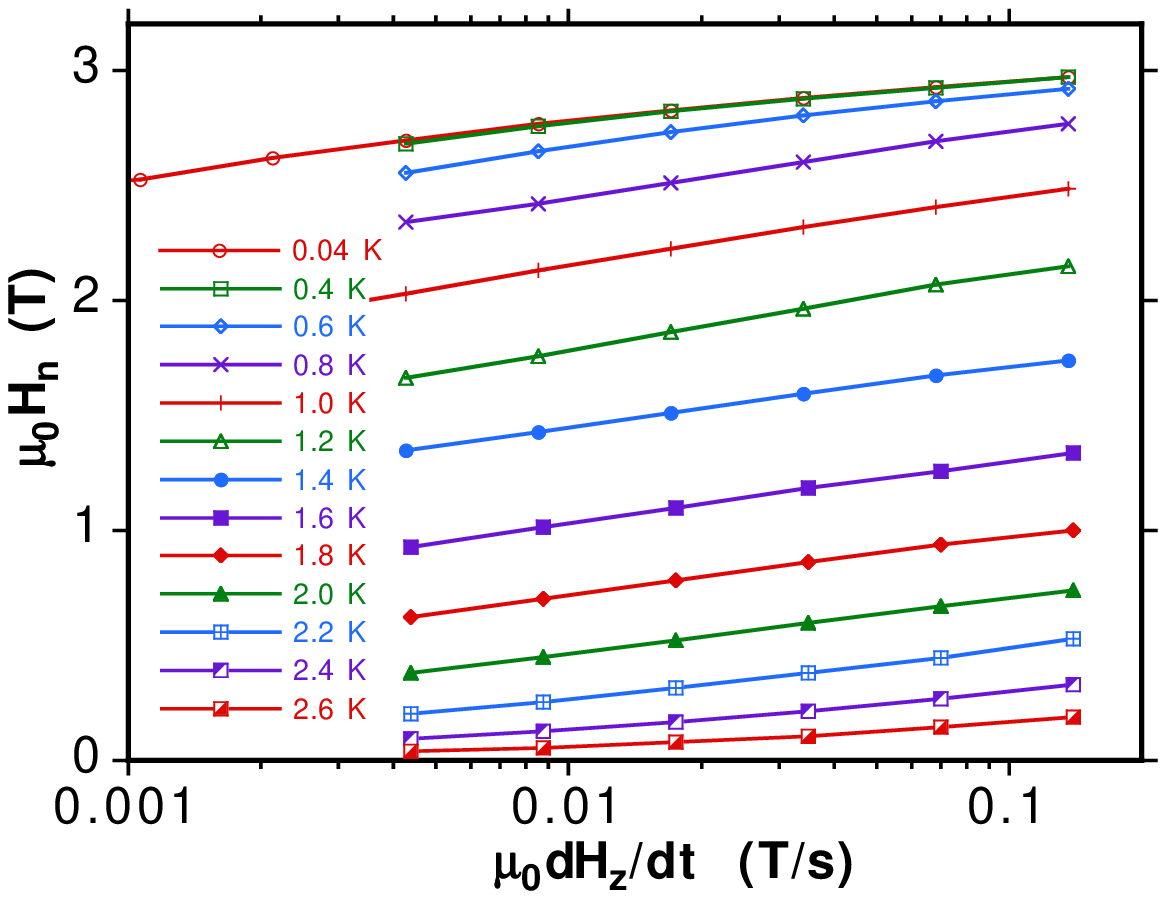}
\caption{(color online) The mean nucleation field $H_{\rm n}$ for the Mn$_{2}$Ni
chain as a function of (a) temperature and (b) field sweep rate.}
\label{Hc_Mn2Ni}
\end{center}
\end{figure}

Typical hysteresis loops are presented in Fig.~\ref{hyst_Mn2Ni}.
The Mn$_{2}$Ni chain displays smooth 
hysteresis loops which are strongly temperature and field sweep rate
dependent. The temperature and field sweep rate dependences
of the coercive fields (called mean nucleation 
fields $H_{\rm n}$ henceforth) were measured and
plotted in Fig~\ref{Hc_Mn2Ni}.
As expected for a thermally
activated process, $H_{\rm n}$ increases
with decreasing temperature $T$ and increasing field sweep
rate $v=dH_z/dt$. Furthermore, all our measurements showed an
almost logarithmic dependence of $H_{\rm n}$ on the field sweep
rate (Fig.~\ref{Hc_Mn2Ni}b).
$H_{\rm n}$ becomes temperature independent below about 0.5 K.

We analyzed the set of $H_{\rm n}(T,v)$ data
with a model of thermally activated nucleation
of magnetization reversal
analogous to that of a magnetic single-domain
particle~\cite{Neel49a,Brown63b,Braun94d,Coffey95}. 
AC and DC relaxation measurements at $\vec{H}=0$
showed that the magnetization reversal below 2.7 K
is dominated by the ends of
the Mn$_2$Ni chain~\cite{Coulon04} and the
dynamics were described by the generalized Glauber model
taking into account chains with finite lengths.
At sufficiently low temperatures and at zero field, 
the energy barrier between the two states of opposite 
magnetization is much too high to observe a reversal. 
However, the barrier can be lowered by applying a magnetic field 
in the opposite direction to that of the chain's magnetization. 
When the applied field is close enough to the 
nucleation field of a domain wall, thermal fluctuations 
are sufficient to allow the system to overcome the nucleation barrier, 
and a domain wall nucleates. Then, due to the applied field, 
the magnetization of the entire chain reverse 
via a domain wall propagation process.
The domain wall nucleation can be 
thermally activated at high temperatures 
or driven by quantum tunneling at low 
temperatures~\cite{Stamp91,Chudnovsky92,Tatara94,Gunther94,Braun97,Shibata00}.

This stochastic nucleation process can be studied via the 
relaxation time method consisting of the measurement of the
probability that the magnetization has not 
reversed after a certain time. In the case
of an assembly of identical and isolated
spin chains, it corresponds to measurements
of the time dependence of magnetization.
The probability 
that the magnetization has not 
reversed after a time $t$ is given by:
\begin{equation}
	P(t) = e^{-t/\tau}
\label{eq_P_t}
\end{equation}
and $\tau$ can be expressed by an 
Arrhenius law of the form:
\begin{equation}
	\tau(T,H) = \tau_0 e^{\Delta E(H)/k_{\rm B}T}
\label{eq_tau}
\end{equation}
where $\Delta E(H)$ is the field dependent nucleation energy barrier
and $\tau_0$ is a prefactor which is supposed to be
a constant.
In most cases $\Delta E(H)$ can be approximated by:
\begin{equation}
	\Delta E(H) \approx E_0 \left(1 - h\right)^\alpha
\label{eq_E}
\end{equation}
where $h = H/H_{\rm n}^0$,
$H_{\rm n}^0$ is the nucleation field at zero temperature,
$E_0$ is roughly the nucleation barrier height at zero applied field, and
$\alpha$ is a constant of the order of unity 
(for most cases $1.5 \leq \alpha \leq 2$~\cite{Victora89}).

There are two limiting cases:
(i) for $D << J$, the nucleation energy barrier
can be calculated exactly~\cite{Braun94d,Braun99,Hinzke00}:
\begin{equation}
	\Delta E(H) = 2k\sqrt{2JD}({\rm tanh}R-hR)
\label{eq_Braun}
\end{equation}
where $R={\rm arcosh}(\sqrt{1/h})$ 
and $k = 1$ or 2 for the nucleation at the ends 
or insight the chain, respectively.
For $h=0$, this energy is the well-known energy
of one or two domain walls, respectively.
For $h \longrightarrow 1$, $\Delta E(H) \approx 4k/3\sqrt{2JD}(1-h)^{3/2}$;
(ii) for $J$ = 0, the spins are decoupled
and we have the case of a Stoner-Wohlfarth particle~\cite{Neel47,St_W48} with
uniaxial anisotropy. When the field is applied along the easy
axis of magnetization, all constants can be determined
analytically~\cite{Neel47,Neel49a}: $\alpha = 2$, $E_0 = KV$,
and the switching field $H_{\rm sw}^0 = 2 K/M_{\rm s}$, where $K$ is the 
uniaxial anisotropy constant, $V$ is the particle
volume, and $M_{\rm s}$ is the saturation magnetization.
For SMMs with dominating uniaxial anisotropy:
$\alpha = 2$, $E_0 = DS^2$,
and $H_{\rm sw}^0 = 2DS/g\mu_0\mu_{\rm B}$.

In our case of the Mn$_2$Ni chain with
$D/k_{\rm B}$ = 2.5 K and $J/k_{\rm B}$ = 1.56 K
we are not aware of an analytical expression
and propose to 
use $\alpha = 2$ because of the arguments developped
in~\cite{Victora89}.

In order to study the field dependence of the relaxation
time $\tau(T,H)$ and to obtain the parameters
of the model,
the decay of magnetization has to be studied at many
applied fields $H$ and temperatures $T$. This is experimentally
very time consuming.
A more convenient method for studying the magnetization decay 
is by ramping the applied field at a 
given rate and measuring 
the mean nucleation field $H_{\rm n}$ which is the field value 
to obtain zero magnetization (coercive field).
$H_{\rm n}$ is then measured as a function of 
the field sweep rate and temperature (Fig~\ref{Hc_Mn2Ni}).
An analogous procedure~\cite{Kurkijarvi72,Gunther94,Garg95} was applied to 
nanoparticles~\cite{WW_PRL97_Co}
and recently to SMMs~\cite{WW_condmat_0509193}.
The mean nucleation field of an
assembly of identical non-interacting SCMs is given by:
\begin{equation}
H_{\rm n}(T,v) \approx H_{\rm n}^0 \left(
	1 - \left\lbrack
	\frac {kT}{E_0} 
	\ln\left(\frac {c}{v}\right)
	\right\rbrack^{1/\alpha}
\right) 
\label{eq_Hc}
\end{equation}
where the field sweeping rate is given by $v = dH_z/dt$; 
$H_{\rm n}^0$ is the nucleation field at zero temperature, and
$c$ depends on the details of the approximations:
$c = H_{\rm n}^0k_{\rm B}T/[\tau_0\alpha E_0(1-H_{\rm n}/H_{\rm 
c}^0)^{\alpha-1}]$ in~\cite{WW_PRL97_Co}, 
$c' = H_{\rm n}^0(E_0/kT)^{1/\alpha}/(\tau_0\alpha)$
in~\cite{Garg95}, and it can be taken constant
when the exact value of $\tau_0$ is not needed.
We applied the three approximations to 
nanoparticles~\cite{WW_PRL97_Co}
and to SMMs~\cite{WW_condmat_0509193} and found that the first approximation gives a
$\tau_0$ which is closest to that extracted from an
Arrhenius plot.

\begin{figure}
\begin{center}
\includegraphics[width=.45\textwidth]{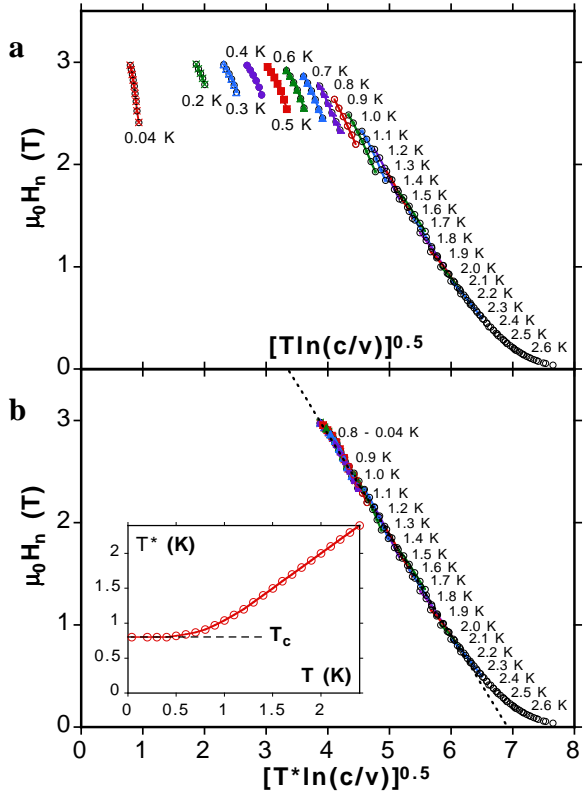}
\caption{(color online) (a) Scaling plot of the mean nucleation field
$H_{\rm n}(T,v)$ for field sweep rates
between 0.001 and 0.14 T/s and several temperatures: 
0.04 K and from 0.2 to 2.6 K in steps of 0.1 K. 
(b) Same data of $H_{\rm n}(T,v)$ and same scales but the
real temperature $T$ is replaced by an effective temperature $T^*$ 
(see inset)
which restores the scaling below 1.1 K.}
\label{scaling_Mn2Ni}
\end{center}
\end{figure}

The validity of Eq.~\ref{eq_Hc} was tested by
plotting the set of $H_{\rm n}(T,v)$ values as a function of
$[Tln(c/v)]^{1/2}$ where 
$c = H_{\rm n}^0k_{\rm B}T/\tau_0 2 E_0(1-H_{\rm n}/H_{\rm n}^0)$.
If the underlying model is sufficient, 
all points should collapse onto one straight line by
choosing the proper values for the constant $\tau_0$. 
We found that the data of $H_{\rm n}(T,v)$ with $T > 1K$ fell on a master curve
provided $\tau_0 = 7.4\times 10^{-9}$ s (Fig.~\ref{scaling_Mn2Ni}).

At lower temperatures, strong deviation from the master
curves are observed. 
In order to investigate the possibility that these 
low-temperature deviations are due to escape 
from the metastable potential well by tunneling, a common method 
for classical models is 
to replace the real temperature $T$ by an effective temperature 
$T^*(T)$ in order to restore the scaling plot~\cite{WW_PRL97_Co,WW_condmat_0509193}. 
In the case of tunneling, $T^*(T)$ should saturate at low temperatures. 
Indeed, the ansatz of $T^*(T)$, as shown in the inset 
of Fig.~\ref{scaling_Mn2Ni}b, can restore unequivocally 
the scaling plot demonstrated by a straight master 
curve (Fig.~\ref{scaling_Mn2Ni}b). 
The flattening of $T^*$ corresponds to a saturation 
of the escape rate, which is a necessary signature of 
tunneling. 
The crossover temperature $T_{\rm c}$ can be defined as 
the temperature where the quantum 
rate equals the thermal one. 
The inset 
of Fig.~\ref{scaling_Mn2Ni}b
gives $T_{\rm c}$ = 0.8 K.
The slope and the intercept of the master curves give 
$E_0/k_{\rm B}$ = 47~K and $\mu_0H_{\rm n}^0$ = 6.95~T.

Several points should be mentioned:
(i) Eq.~\ref{eq_Hc} is not valid for fields which
are close to $H=0$ because the model only takes into account
the nucleation from the metastable to the stable state.
However, close to $H=0$, transitions between both states are possible
leading to a rounding of the master curve at small fields;
(ii) the remaining energy barrier for the tunnel process
at 0.04 K can be calculated
using Eq.~\ref{eq_E}. We found $\sim$30~K for a field sweep rate 
of 0.001~T/s;
(iii) in case of a distribution of nucleation barriers, different
parts of the distribution can be probed by applying the 
method at different $M$ values;
(iv) this method is insensitive to small intermolecular
interactions when $H_{\rm n}$ is much larger than the typical
interaction field; and
(v) the method can be generalized for 2D and 3D networks
of spins. 

In conclusion, the presented low-temperature studies
of the field driven magnetization reversal 
of the Mn$_2$Ni SCM suggest for the 
low temperature region that the 
magnetization reversal starts by a quantum
nucleation of a domain wall followed
by domain wall propagation and reversal
of the magnetization. Further studies
will concern the application of transverse fields
which should enhance the quantum nucleation rate.

This work was supported by the EC-TMR Network 
QuEMolNa (MRTN-CT-2003-504880), CNRS,
Rh${\rm\hat{o}}$ne-Alpes funding,
the Universit\'e Bordeaux 1, and the Conseil R\'egional
dÕAquitaine.

\begin{figure}
\begin{center}
\includegraphics[width=.45\textwidth]{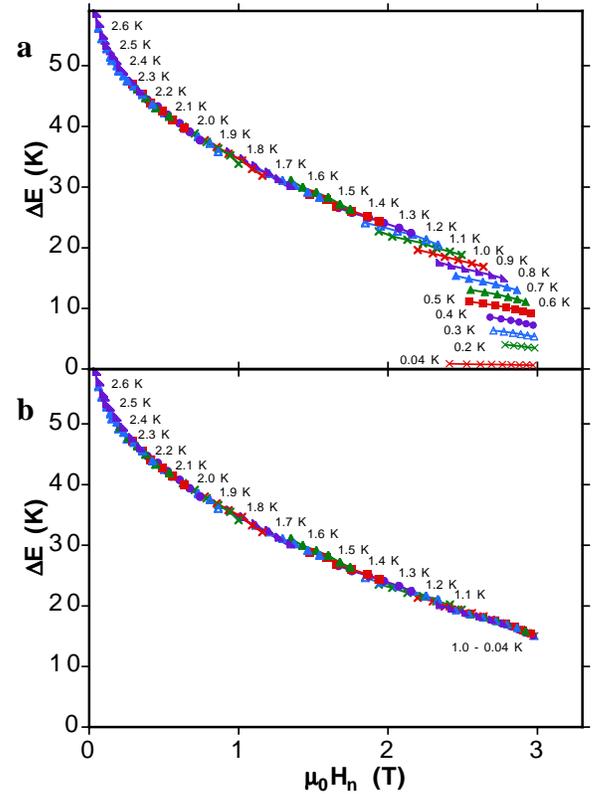}
\caption{(color online) (a) Field dependence of the
energy barrier of the Mn$_2$Ni chain obtained from Eq.~\ref{eq_barrier}
and the set of $H_{\rm c}(T,v)$ data from Fig.~\ref{scaling_Mn2Ni}. 
(b) Same data of $H_{\rm c}(T,v)$ but the
real temperature $T$ is replaced by an effective temperature $T^*$ 
(see inset of Fig.~\ref{scaling_Mn2Ni}b).}
\label{barrier_Mn2Ni}
\end{center}
\end{figure}

\section{Annex}

The probability density of reversal of a stochastic process is
\begin{equation}
    -\frac{dP}{dt} = \frac{1}{\tau}P
\label{eq_density}
\end{equation}
and the maximum of the probability density can be
derived from
\begin{equation}
    \frac{d^2P}{dt^2} =  
    \frac{P}{\tau^2}\left(1+\frac{d\tau}{dt}\right) = 0
\label{eq_d2P}
\end{equation}
This gives the following general result for the 
maximum of the probability density
\begin{equation}
    \frac{d\tau}{dt} = -1
\label{eq_max_density}
\end{equation}
The application of the result to Eq.~\ref{eq_tau} leads to
\begin{equation}
    \Delta E(H) = k_{\rm B}T{\rm ln}\left(\frac{k_{\rm 
    B}T}{\tau_0\frac{dE}{dH}\frac{dH}{dt}}\right)
\label{eq_barrier}
\end{equation}
Using Eqs.~\ref{eq_E} and~\ref{eq_barrier}, we find  Eq.~\ref{eq_Hc}.

Eq.~\ref{eq_barrier} can be used to plot directly the
field dependence of the energy barrier (Fig.~\ref{barrier_Mn2Ni})


\end{document}